\documentstyle[aasms4,12pt] {article}
\begin{document}

\title{The Detection of Massive Molecular Complexes\\
in the Ring Galaxy System Arp 143}
\vskip 0.25in

\author{James L. Higdon\footnotemark[1]}
\affil{CSIRO/Australia Telescope National Facility\\
       Paul Wild Observatory\\
       Locked Bag 194, Narrabri, NSW 2390\\
       E-mail: jhigdon@atnf.csiro.au}

\author{Richard J. Rand}
\affil{Department of Physics and Astronomy\\
       University of New Mexico\\
       800 Yale Blvd, NE\\
       Albuquerque, NM  87131\\
       E-mail: rjr@gromit.phys.unm.edu}

\begin{center}
and
\end{center}

\author{Steven D. Lord}
\affil{IPAC/Caltech, MS 100-22 \\
       Pasadena, CA 91125\\
       E-mail: lord@ipac.caltech.edu}

\footnotetext [1] {ATNF Fellow}

\vfil \eject

\begin{abstract}
We have imaged the kpc scale distribution of
$^{12}$CO(J=1-0) emission in the ring
galaxy system Arp 143 (NGC 2444/2445) using the OVRO millimeter
array.  We find two giant molecular complexes in the ring component 
(NGC 2445) and a bright central source.  The ring complexes
represent 20$-$60$\%$ of the detected M$_{H_{2}}$, depending on the 
relative I$_{CO}$--$N_{H_2}$ for the ring and nucleus.
Their individual H$_{2}$ masses and surface
densities ($\Sigma_{H_2}$) exceed typical spiral arm GMAs 
regardless of the conversion factor.  Both are
associated with a 6 kpc ridge of peak $\Sigma_{HI}$
and massive star formation (MSF) activity.  H$\alpha$ imaging 
shows a patchy ring of HII regions situated along 
the outer edge of the HI ring.
The kinematics of the HI ring show clear signs of expansion.
A simple rotating-expanding ring model (V$_{exp}$ = 118 $\pm$ 30
km s$^{-1}$) fits the data reasonably well, implying a ring
age of $\gtrsim$60 $\pm$ 15 Myrs.  NGC 2445's ring is
able to promptly form very large molecular complexes in a metal
poor ISM and trigger MSF.

Nearly eighty per cent of the detected $^{12}$CO(1-0) flux
originates in a resolved central source 
that is slightly offset from NGC 2445's starburst nucleus.
We find an ordered velocity field in this component.
Assuming an inclined disk, we argue that it is dynamically stable. 
The central $\Sigma_{H_{2}}$ (910 M$_{\odot}$ pc$^{-2}$) 
significantly exceeds $\Sigma_{H_{2}}$ commonly found in
normal spirals, but is much smaller than values 
derived in similar sized regions of IR luminous galaxies.
The nuclear H$_{2}$ may be the result of a previous encounter
with NGC 2444.  $^{12}$CO(1-0) emission in ring galaxies may be
dominated by the nucleus, which could bias the 
interpretation of single-dish measurements.
\end{abstract}


\vfill \eject

\section{Introduction}
HI studies of such archetype 
ring galaxies as the Cartwheel and AM0644-741 show that
$\sim$90$\%$ of the neutral atomic ISM is concentrated in the narrow 
orbit crowded rings for at least 40 Myrs (Higdon 1996; Higdon, Wallin, 
Staveley-Smith, $\&$ Lord 1997).  This is thought to create an environment 
conducive to the formation of massive cloud complexes and
the triggering  of massive star formation (MSF)
through elevated collision rates.
However, we know little of the molecular ISM in these systems, 
which is unfortunate since massive
stars in normal galaxies form exclusively in giant molecular 
clouds and associations (GMAs).  Investigations of the molecular component 
of ring galaxies through the prominent 115 GHz $^{12}$CO(J=1-0) 
transition has proven difficult due to their $\sim$1$'$ angular size,
which is comparable to the beams of most single-dish mm-wave
telescopes, and typically low $^{12}$CO(1-0) fluxes (Horellou et al.
1995; Higdon, Lord, $\&$ Rand 1997).  This has led to speculation
that MSF in ring galaxies occurs in a primarily {\em atomic} ISM (Higdon 1996).
In this Letter we present the first aperture synthesis 
$^{12}$CO(J=1-0) observations of a putative ring galaxy system -- Arp 143 
(NGC 2444/2445), which possesses the largest 115 GHz line flux in
our ring galaxy survey.  Complementary
HI and H$\alpha$ data are also presented to explore the relationship
between the neutral ISM and MSF on similar spatial
scales.  We will show that NGC 2445 is indeed a young ring galaxy, whose
ring is forming unusually large GMAs in a metal poor ISM and triggering
MSF.

Optical images of NGC 2445, the Southern component in Arp 143,
show a compact nucleus surrounded by a number of blue knots which are
in fact low metallicity\footnotemark[2]
HII regions (Burbidge $\&$ Burbidge 1959; Jeske 1986). 
\footnotetext [2] {12 + log[O/H] = 8.56-8.77, Jeske (1986).}
Appleton, Schombert, $\&$ Robson (1992) found that most of the knots have
optical/NIR colors consistent with very young (Age $\lesssim$ 10 Myrs) stellar
populations, and are located along the outer edge of an 
HI crescent, the distribution expected in ring galaxies created 
in an off-centered ``intruder'' passage.  They proposed that NGC 2445
represented a young ``pre-starburst'' ring galaxy.  The HI crescent's 
kinematics were not discussed, and no ring expansion - the {\em sine qua non} 
for a collisional ring galaxy - was reported.  We presented 
$^{12}$CO(1-0) single-dish results for Arp 143 in
Higdon et al. (1995) and noted an enhancement in line emission Northwest 
of NGC 2445's nucleus.  We proposed that this was due to molecular gas in the
HI crescent.  

\section{Observations and Data Reduction}
$^{12}$CO(J=1-0) observations of Arp 143 were made using the six-element 
Owens Valley Radio Observatory millimeter-wave interferometer
on 2 $\&$ 20 May 1995 in the compact ``A'' configuration 
(20.8--200 m baseline range).  Each 10.4 m antenna was 
equipped with an SIS receiver cooled to 4 K. Single sideband T$_{sys}$
were between 300$-$500 K at 113 GHz.  The correlator was 
configured for 120 channels separated by 2 MHz (5.2 km s$^{-1}$),
giving a usable velocity range of 660 km s$^{-1}$.  The array tracked a
position near NGC 2445's optical nucleus for
two 12-hour runs. Time dependent amplitude and phase variations
were monitored through frequent observations of QSO 0642+449,
while 3C 273 was used to define the flux scale and correlator bandpass
shape.  We estimate the flux calibration uncertainty to be 10$\%$.
The UV data were processed
using the OVRO reduction package MMA (Scoville et al. 1992).  Channel
maps were made using both natural ($\theta_{FWHM}$ = 5$''$, 1$\sigma$ = 
12 mJy beam$^{-1}$, ${\small\Delta}$V = 10.4 km $^{-1}$)
and uniform ($\theta_{FWHM}$ = 3.9$''$, 1$\sigma$ = 14 mJy beam$^{-1}$,
${\small\Delta}$V = 20.8 km $^{-1}$) weighting in AIPS, and CLEANed.
The naturally weighted maps were further convolved
to increase sensitivity ($\theta_{FWHM}$ = 8$''$, 1$\sigma$ = 10 mJy 
beam$^{-1}$).  At the assumed distance of 40 Mpc (H$_{\circ}$
= 100 km s$^{-1}$ Mpc$^{-1}$) this corresponds to
a spatial resolution of 1.6 kpc.  Integrated $^{12}$CO(1-0) maps were 
made by summing emission that exceeded the 3$\sigma$ 
map noise, showed spatial and velocity continuity over at 
least 3 consecutive channels, and was within 150 km s$^{-1}$ of
the local HI velocities. The final maps were corrected for
primary beam attenuation. No continuum emission was detected.

Arp 143 was observed with the Very Large Array$\footnotemark[3]$ on
27 December 1995 in B-configuration to image HI emission with
velocity and spatial resolution comparable with the OVRO data.  Details
of those observations will be presented elsewhere.  Naturally
weighted channel maps ($\theta_{FWHM}$ = 5.4$''$, 1$\sigma$
= 0.3 mJy beam$^{-1}$, $\small\Delta$V = 10.6 km s$^{-1}$) were used
to construct HI moment maps using routines in AIPS.  IDL programs
were used to analyze the HI kinematics.
$\footnotetext [3] {The National Radio Astronomy Observatory is a facility
of the National Science Foundation operated under
cooperative agreement by Associated Universities, Inc.}$

We obtained H$\alpha$ + [N II] and Johnson B-band CCD images
of Arp 143 with the McDonald Observatory 0.76 m telescope 
on 10-12 March 1992 using the TI2 CCD located at the Cassegrain focus
(0.58$''$ pix$^{-1}$).  Line emission
was isolated using an 80 {\AA}  FWHM filter centered at 6650 {\AA}, while
line-free continuum was mapped using a wider (120 {\AA} FWHM)
filter centered blueward of H$\alpha$.  
The CCD reduction was routine and performed in IRAF.  The line
map was calibrated using extinction corrected
HII region line fluxes from Jeske (1986),
and is believed accurate to $\sim$25$\%$, except in the nucleus
where the images saturate.   The H$\alpha$ image was smoothed 
($\theta_{FWHM}$ = 5$''$) to improve sensitivity and better match the
$^{12}$CO(1-0) and HI data. 
Measured field star positions were used to define an astronomical
coordinate system in the optical images to an accuracy of 
1$''$ (Benedict $\&$ Shelus 1978).

\section{Results}

\subsection{The Neutral ISM of NGC 2445}

Contours of integrated $^{12}$CO(1-0) and HI line intensity are shown
superposed on the Arp 143 B-band image
in Figures 1a $\&$ b.  We find (1.25 $\pm$ 0.08) $\times$ 10$^{9}$ M$_{\odot}$
of atomic hydrogen concentrated into an $\sim$15 kpc diameter ring
in NGC 2445's disk.  Though we detect only 25$\%$ of the single-dish 
21cm flux (Shostak 1978), our data provides a high resolution view of
the HI ring, the most conspicuous component in Appleton et al.'s (1992) 
C- and D- array VLA maps.  The highest $\Sigma_{HI}$ is found in a  
quasi-linear ridge West of the ring's optical nucleus (30-60 M$_{\odot}$ 
pc$^{-2}$).  Surface densities between 3 and 20 M$_{\odot}$ 
pc$^{-2}$ typify the rest of the HI crescent.
We do not detect HI in NGC 2445's nucleus or in NGC 2444.
By contrast, $^{12}$CO(1-0) emission in NGC 2445 is dominated
by a slightly resolved nuclear source, plus two large 
complexes (``a'' $\&$ ``b'') $\sim$20$''$ to the West
$\&$ Northwest.  Both are associated with the ring's high $\Sigma_{HI}$
ridge.  A third elongated source (``c'') extends Southeast 
from the nucleus into a low $\Sigma_{HI}$ region.
Their emission properties are listed in Table 1.  Altogether, 
we measure a $^{12}$CO(1-0) flux integral of 36.7 $\pm$ 2.1 
Jy km s$^{-1}$ within the OVRO primary beam, or 33$\%$ of the
single-dish value (Higdon et al. 1995).  Adopting a Galactic 
I$_{CO}$-N$_{H_2}$ conversion factor (X$_{Gal}$, Bloemen et al. 1986) 
leads to a total H$_{2}$ mass  of (1.2 $\pm$ 0.1) $\times$ 10$^{9}$ M$_{\odot}$.  But 
given the LMC-like metallicity of the ring, use of an LMC
conversion factor (X$_{LMC}$, Cohen et al. 1988) may be more appropriate
for ``a'' and ``b'', in which case the total M$_{H_2}$ becomes
(2.2 $\pm$ 0.2) $\times$ 10$^{9}$ M$_{\odot}$.  The two ring GMAs would then
represent 60$\%$ of the total molecular gas detected by the
array.  We will dodge the uncertain conversion factor issue 
by listing a range of M$_{H_2}$ and 
$\Sigma_{H_2}$ corresponding to the two choices of X.

The two molecular complexes of the ring contribute
19$\%$ of the observed $^{12}$CO(1-0) emission.  Their individual 
H$_{2}$ masses are 1.6--9.7 $\times$ 10$^{8}$ M$_{\odot}$ (``a'')
and 0.4--2.5 $\times$ 10$^{8}$ M$_{\odot}$ (``b'').  Compared with the
GMAs in M100 and M51, mapped with 700 pc resolution by 
Rand (1995, 1993), NGC 2445's
complexes are more massive with a comparable range in $\Sigma_{H_2}$
assuming X$_{Gal}$ (Table 1).  NGC 2445's $\Sigma_{H_{2}}$'s are 
no doubt lower limits due to the 1.6 kpc naturally weighted synthesized beam.
At higher resolution they might break up into less massive GMAs as well.
However, if X$_{LMC}$ applies, their M$_{H_2}$ and  $\Sigma_{H_{2}}$ 
significantly exceed the GMAs in M100 $\&$ M51, even when
making allowances for differing resolutions.  Both ``a'' and ``b'' 
are found in high $\Sigma_{HI}$ regions ($\sim$50-60 M$_{\odot}$ 
pc$^{-2}$), though HI and $^{12}$CO(1-0) peaks do not coincide.  
Regions where $\Sigma_{HI}$ $>$ 20 M$_{\odot}$ pc$^{-2}$ and where no
$^{12}$CO(1-0) emission is detected are found within the OVRO primary 
beam.  H$_{2}$ may of course be
present here as our map noise sets a point source M$_{H_2}$
detection limit of 1.8 $\times$ 10$^{7}$ M$_{\odot}$ (3$\sigma$)
for $\Delta$V of 30 km s$^{-1}$ and X$_{Gal}$.
Significant differences between the $^{12}$CO(1-0) and HI velocities
were found in the ring, with $\Delta$V$_{CO-HI}$ equal to -38 
km s$^{-1}$ for ``a'' and +40 km s$^{-1}$ for ``b''.  
Both are located in regions of
large HI velocity dispersion or streaming, so it is difficult
to attribute the offsets to an obvious systematic effect.

More than 80$\%$ of the total $^{12}$CO(1-0) flux, or (7.1 $\pm$ 1.6)
$\times$ 10$^{8}$ M$_{\odot}$ of H$_{2}$ (X$_{Gal}$), is emitted 
by NGC 2445's nucleus.
The $^{12}$CO(1-0) peak is displaced  5$''$ (0.8 kpc) Southwest
of the optical center.
A uniformly weighted $^{12}$CO(1-0) map of this region
is presented in Figure 2a, and 
shows the emission to be clearly elongated along a 
position angle of 71$^{\circ}$,  with a minor/major axis ratio of 
$\sim$2/3.  NGC 2445's peak $\Sigma_{H_{2}}$ 
(910 M$_{\odot}$ pc$^{-2}$, X$_{Gal}$) is
significantly larger than $\Sigma_{H_{2}}$ inferred in the
nuclei of normal spirals ($\overline{\Sigma}_{H_2}$ = 190 M$_{\odot}$
pc$^{-2}$ in 8 Virgo spirals, with an 88 M$_{\odot}$
pc$^{-2}$ rms, Canzian 1990) and some mergers 
(e.g., NGC 4038/39, Stanford et al. 1990).  However, NGC 2445's central
M$_{H_2}$ and $\Sigma_{H_{2}}$ are far outclassed by the 
centers of IR luminous galaxies like Mrk 273 ($\Sigma_{H_{2}}$ =
3.8 $\times$ 10$^{4}$ M$_{\odot}$ pc$^{-2}$, Yun $\&$
Scoville 1995).  These comparisons are summarized in Table 1.

The velocity field of the nuclear  $^{12}$CO(1-0)
emission (Figure 2b) 
shows ordered motions suggesting a disk geometry. 
We estimated its dynamical mass assuming a radius of 
490 pc and 45$^{\circ}$ $\pm$ 10$^{\circ}$ inclination to be 
\begin{equation}
M_{dyn} = \frac{R\;V^{2}_{1/2}}{G \;sin^{2}{\it i}} = (3.3 \pm 1.3) \times 10^{9}
M_{\odot}.
\end{equation}
The ratio of molecular to dynamical mass M$_{H_{2}}$/M$_{dyn}$
is 0.2 $\pm$ 0.1 using X$_{Gal}$,
and is consistent with a system in equilibrium given the
uncertainties. This is also similar to values determined for the 
nuclei of normal spirals, as opposed to the 
molecular masses and M$_{H_{2}}$/M$_{dyn}$ typically
found in IR luminous galaxies (Yun et al. 1994).
Finally, we tentatively identify the elongated source ``c''
with a faint dust lane visible in the Arp Atlas (1966) photograph.

\subsection{Massive Star Formation and the ISM in NGC 2445}

Figure 1c shows the distribution of HII regions in Arp 143 with 
L$_{H\alpha}$ between 0.6 - 15. $\times$ 10$^{39}$ erg s$^{-1}$.  
All are found in NGC 2445.  Apart from the nucleus, MSF is
concentrated along a patchy ellipse having the same general
shape as the HI crescent in Figure 1b.   The 3 most luminous
HII regions form an $\sim$6 kpc complex associated with both 
the GMAs ``a'' $\&$ ``b'' and the high $\Sigma_{HI}$ ridge 
of the ring.  We estimate the ring's SFR to be 0.5 M$_{\odot}$ 
yr$^{-1}$ based on its L$_{H\alpha}$, which is 25$\%$ of the 
nuclear SFR (Kennicutt 1983, Jeske 1986).  This is only 25$\%$ of
the average SFR for four similarly sized ring galaxies in Marston
$\&$ Appleton's (1995) survey, and less than 1$\%$ of the 
Cartwheel's (Higdon 1995).  The global relationship 
between HI, $^{12}$CO(1-0), and H$\alpha$ emission is illustrated in 
Figure 1d.    Note that with one exception HII regions are
situated along the outer edge of the $\Sigma_{HI}$ $>$ 10 
M$_{\odot}$ pc$^{-2}$
ring.   This is most pronounced in the South and East, but is also true in
the HI ridge West of the nucleus.  We detect 
no HII regions with L$_{H\alpha}$ above 3 $\times$ 10$^{38}$ 
erg s$^{-1}$ (3$\sigma$) in NGC 2445 apart from the nucleus and ring,
or in NGC 2444.

\subsection{Kinematics of the HI Ring}

In Figure 3 we show a radial velocity-position angle diagram 
for the HI ring.  The data represent averages within 10$^{\circ}$
segments of a 6$''$ wide annulus fit to the HI distribution.
The dashed line shows the best fit to a {\em non-expanding} rotating 
circular ring model. Strong non-circular motions are seen along the minor axis 
on both sides of the ring (p.a. = 40-130$^{\circ}$ and 250-340$^{\circ}$)
which are most easily explained as expansion.  The solid line shows 
a least-squares fit of a 
rotating-expanding circular ring model, from which we derive
V$_{sys}$ = 4000 $\pm$ 10 km s$^{-1}$,  V$_{rot}$ = 214 $\pm$ 25 km s$^{-1}$,
and V$_{exp}$ = 118 $\pm$ 30 km s$^{-1}$.   The rotating-expanding ring
is not a perfect fit.  This may reflect non-planar motions, tidal interaction
with NGC 2444, or departures from circularity.  Note that the largest
discrepancy occurs near p.a. = 180$^{\circ}$, which is where the
HI ring splits in two. Nevertheless, expansion
is clearly indicated by the data.   From the ring's radius and
V$_{exp}$ we estimate its age to be $\sim$60 $\pm$ 15 Myrs, or 
20$\%$ of the Cartwheel's (Higdon 1996).

\section{Discussion}

These observations show that NGC 2445's ring contains large
scale concentrations of molecular gas.  At 1.6 kpc resolution,
their M$_{H_2}$ and $\Sigma_{H_2}$ appear larger
than those typically found in spiral arms, especially if
X$_{LMC}$ is adopted.
Their close association with the highest $\Sigma_{HI}$ of the ring
suggests their formation in the orbit crowded ring density wave
through agglomeration or gravitational instability.   The ring
is clearly capable of forming large molecular masses out of a
metal poor ISM.  Similarly, the concentration of luminous 
metal poor HII regions along the gas ring's outer edge, 
and the close association between peak
$\Sigma_{HI}$, $\Sigma_{H_2}$, and $\Sigma_{H\alpha}$ implies 
direct MSF triggering, rather than the simple rearrangement of
pre-existing star forming regions (Higdon 1995).  

We find the morphology of NGC 2445's neutral and ionized gas to be in
excellent agreement with model ring galaxies formed in off-centered
collisions (cf. Appleton $\&$ Struck 1987).
Indeed, the absence of a stellar bar makes it very unlikely
that the gas ring is a resonance phenomenon.
We found strong evidence of expansion in the HI ring 
implying that the ring
was formed $\gtrsim$60 $\pm$ 15 Myrs ago.
Together with the ring's low metallicity and $\sim$10 Myr
old clusters, the large M$_{H_2}$ and $\Sigma_{H_2}$ for complexes
``a'' and ``b'', and a low SFR relative to other rings,
the body of evidence points to NGC 2445 being a young ring galaxy.

The origin of the nuclear $^{12}$CO(1-0) emission is less certain.
The high $\Sigma_{H_2}$ relative to normal spirals coupled with
the nuclear starburst suggest a tidal origin.  However the
central $\Sigma_{H_2}$ is clearly not in the same class as IR luminous
galaxies.  Published ring galaxy simulations do produce large
mass buildups in central regions, but not until the ring galaxy
is highly evolved (e.g., Appleton $\&$ Struck 1987,
Mihos $\&$ Hernquist 1994).
Appleton et al. (1987) found a narrow
HI plume extending $\gtrsim$100 kpc from Arp 143.  Such a
large structure should be much older than the $\sim$60 Myr
old ring, implying multiple interactions between
NGC 2444 and 2445.  It is conceivable that
a large impact parameter encounter generated the long HI
tidal tail, while at the same time funneling gas to the center
of NGC 2445.  This lowered the total orbital energy
sufficiently to bring NGC 2444 through the disk of
NGC 2445 $\sim$60 Myrs ago, forming the ring.
Recent models suggest that the nuclear gas would
be able to survive NGC 2444's ``intrusion'', especially
if the latter were gas poor (Struck 1997).   The next
encounter will likely lead to a merger of the two galaxies.
These issues should be explored further with detailed
numerical simulations.

Regardless of its origin, there is evidence that
nuclear components may dominate
the molecular emission of ring galaxies.
Of the 6 ring galaxies large enough to be observed with
multiple pointings by Horellou et al. (1995), $^{12}$CO(1-0)
emission peaked at the nucleus in each case.  
If this is true in general,  comparisons between global H$_{2}$
properties derived using low resolution single-dish
measurements and optical H$\alpha$
or B-band fluxes may not be very meaningful, since young 
stars are highly concentrated in the outer rings.
Unless emission from
the nucleus and rings can be distinguished, the analysis may
be subject to large systematic errors.   $^{12}$CO(1-0)
interferometric studies of other ring galaxies are
clearly needed.  However, the best examples of this class
such as the Cartwheel will require the next generation 
of millimeter-wave arrays.

We wish to thank Curt Struck for a number of enlightening discussions,
Philip Appleton for the use of his H$\alpha$ filter, and B. Canzian
for access to some of his unpublished results on Virgo Cluster spirals.
A special word of thanks is also extended to the staffs at
the Owens Valley Radio Observatory,
McDonald Observatory, and NRAO for their help and support.  We
also wish to thank an anonymous referee for suggesting a
number of improvements.  This research has made use of the NASA/IPAC 
Extragalactic Database (NED) which is operated by the Jet 
Propulsion Laboratory, California
Institute of Technology, under contract with the National Aeronautics
and Space Administration.
  
\vfill \eject

\vfill \eject

\begin{center}
{\bf Figure Captions}
\end{center}

Figure 1. $-$ (a) Contours of integrated $^{12}$CO(1-0)
emission (natural weighting, $\theta_{FWHM}$ = 8$''$)
superposed on a B-band CCD image of the Arp 143 system.
The dashed circle indicates the OVRO primary beam FWHM.
The levels are logarithmic and correspond to 0.45, 0.79,
1.4, 2.5, 4.5, 7.9, 14.1, $\&$ 25.1 Jy km s$^{-1}$ beam$^{-1}$.
(b) Contours of $\Sigma_{HI}$ from the
VLA B-array data ($\theta_{FWHM}$ =
5.4$''$). The levels range from 
10 to 50 M$_{\odot}$ pc$^{-2}$ in steps of 10 M$_{\odot}$ pc$^{-2}$.
Crosses mark the positions of HII region centers.
(c) Contours of $\Sigma_{H\alpha}$ in Arp 143.  
Ten logarithmically spaced levels between 7.4 and 340. 
$\times$ 10$^{-17}$ erg s$^{-1}$ cm$^{-2}$ arcsec$^{-2}$ are shown.
(d) Composite showing relative
distribution of $\Sigma_{HI}$ (grey scale), $\Sigma_{CO}$ (contours),
and HII regions (crosses).

Figure 2. $-$ Molecular gas in NGC 2445's nucleus using the
uniformly weighted OVRO data ($\theta_{FWHM}$ = 3.9$''$). (a)
Contours of integrated $^{12}$CO(1-0) emission from 2.3 to 20.7 
Jy km s$^{-1}$ beam$^{-1}$
in steps of 2.3 Jy km s$^{-1}$ beam$^{-1}$.
The cross marks the center of the optical nucleus.
(b) Flux-weighted velocity field.  The contours
are labeled in units of km s$^{-1}$ (optical-Heliocentric).

Figure 3. $-$ Radial velocity-position angle diagram for the HI
ring shown in Figure 1b.  Position angle increases counterclockwise
from the Eastern major axis line of nodes.
The solid line represents a least-squares
fit to an inclined circular, rotating, and expanding ring model,
while the dashed line corresponds to the same model without 
expansion.  The rotating-expanding model gives V$_{sys}$ =
4000 $\pm$ 10 km s$^{-1}$, V$_{rot}$ = 214 $\pm$ 25
km s$^{-1}$, and V$_{exp}$ = 118 $\pm$ 30 km s$^{-1}$.


\begin{references}

\reference{ASM} Appleton, P. N., $\&$ Struck, C.  1987, \apj, 318, 103

\reference{A92} Appleton, P. N., Schombert, J. M., $\&$ Robson, E. I.
1992, \apj, 385, 491

\reference{A87} Appleton, P. N., Ghigo, F. G., Van Gorkom, J. H.,
Schombert, J. M., $\&$ Struck, C.  1987, Nature, 330, 140

\reference{A66} Arp, H. J.  1966, \apjs, 14, 1

\reference{BS} Benedict, G. F., $\&$ Shelus, P. J.  1978, Applications of
Automated Inventory Techniques to Astrometry, in Proceed. I. A. U.
Colloquium No. 48, 109

\reference{B86} Bloemen, J. B., Strong, A. W., Blitz, L., Cohen, R. S.,
Dame, T. M., Grabelsky, D. A., Hermsen, W., Lebrun, F., Mayer-Hasselwander,
H. A., $\&$ Thaddeus, P.  1986, \aap, 154, 25

\reference{jabba} Burbidge, E. M., $\&$ Burbidge, G. R. 1959, \apj, 10, 12

\reference{Can} Canzian, B.  1990, Ph. D. Thesis, California Institute of
Technology

\reference{Cohen} Cohen, R. S., Dame, T. M., Garay, G., Montani, J.,
Rubio, M., $\&$ Thaddeus, P. 1988, \apj, 331, L95

\reference{H95Ha} Higdon, J. L.  1995, \apj, 455, 524

\reference{H95HI} Higdon, J. L.  1996, \apj, 467, 241


\reference{H95} Higdon, J. L., Smith, B. J., Lord, S. D., $\&$
Rand, R. J.  1995, \apj, 438, L79

\reference{HLR} Higdon, J. L., Lord, S. D., $\&$ Rand, R. J.  1997,
in preparation

\reference{HWSSL} Higdon, J. L., Wallin, J. F., Staveley-Smith, L., $\&$
Lord, S. D.  1997, in preparation

\reference{Hor} Horellou, C., Casoli, F., Combes, F., $\&$
Dupraz, C.  1995, \aap, 298, 743

\reference{Jes} Jeske, N. A. 1986, Ph. D. Thesis, University of California,
Berkeley

\reference{K83} Kennicutt, R. A.  1983, \apj, 272, 54


\reference{MA} Marston, A., $\&$ Appleton, P. N.  1995, \aj, 109, 1002

\reference{MiHe} Mihos, J. C., $\&$ Hernquist, L. 1994, \apj, 473, 611 

\reference{R93} Rand, R. J.  1993, \apj, 404, 593

\reference{R95} Rand, R. J.  1995, \aj, 109, 2444

\reference{SnM} Sanders, D. B., $\&$ Mirabel, I. F.  1996, \araa, 34, 749

\reference{mma} Scoville, N. Z., Carlstrom, J. C., Chandler, C. J.,
Phillips, J. A., Scott, S. L., Tilanus, R. P., $\&$ Wang, Z.  1992,
PASP, 105, 1482

\reference{shos} Shostak, G. S.  1978, \aap, 68, 321

\reference{Stan} Stanford, S. A., Sargent, A., Sanders, D. B.,
$\&$ Scoville, N. Z.  1990, \apj, 349, 492

\reference{CSM} Struck, C.  1997, private communication

\reference{Yun} Yun, M. S., Scoville, N. Z., $\&$ Knop, R. A.  1994,
\apj, 430, L109

\reference{Yun95} Yun, M. S., $\&$ Scoville, N. Z.  1995, \apj, 451, L45

\end{references}
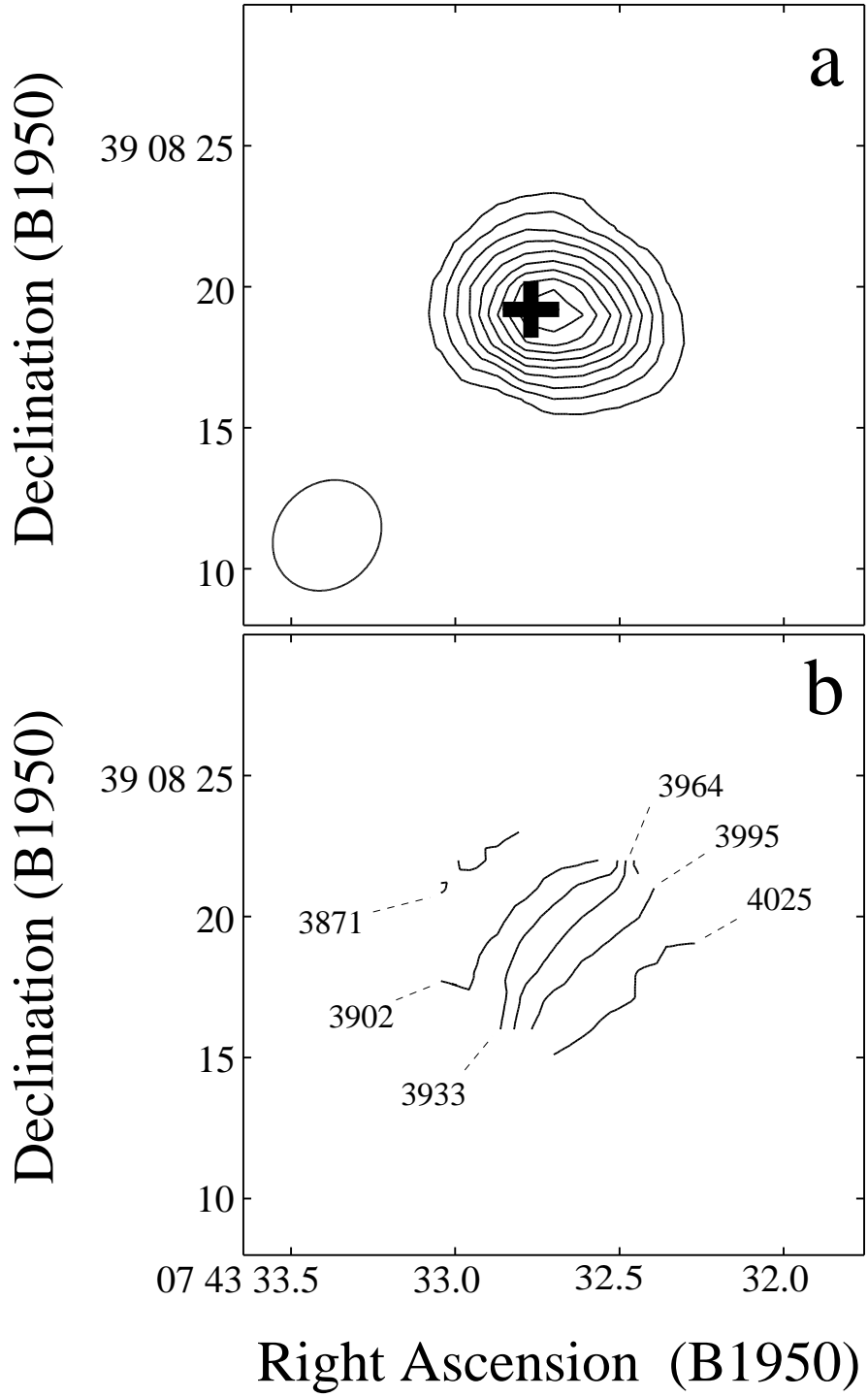
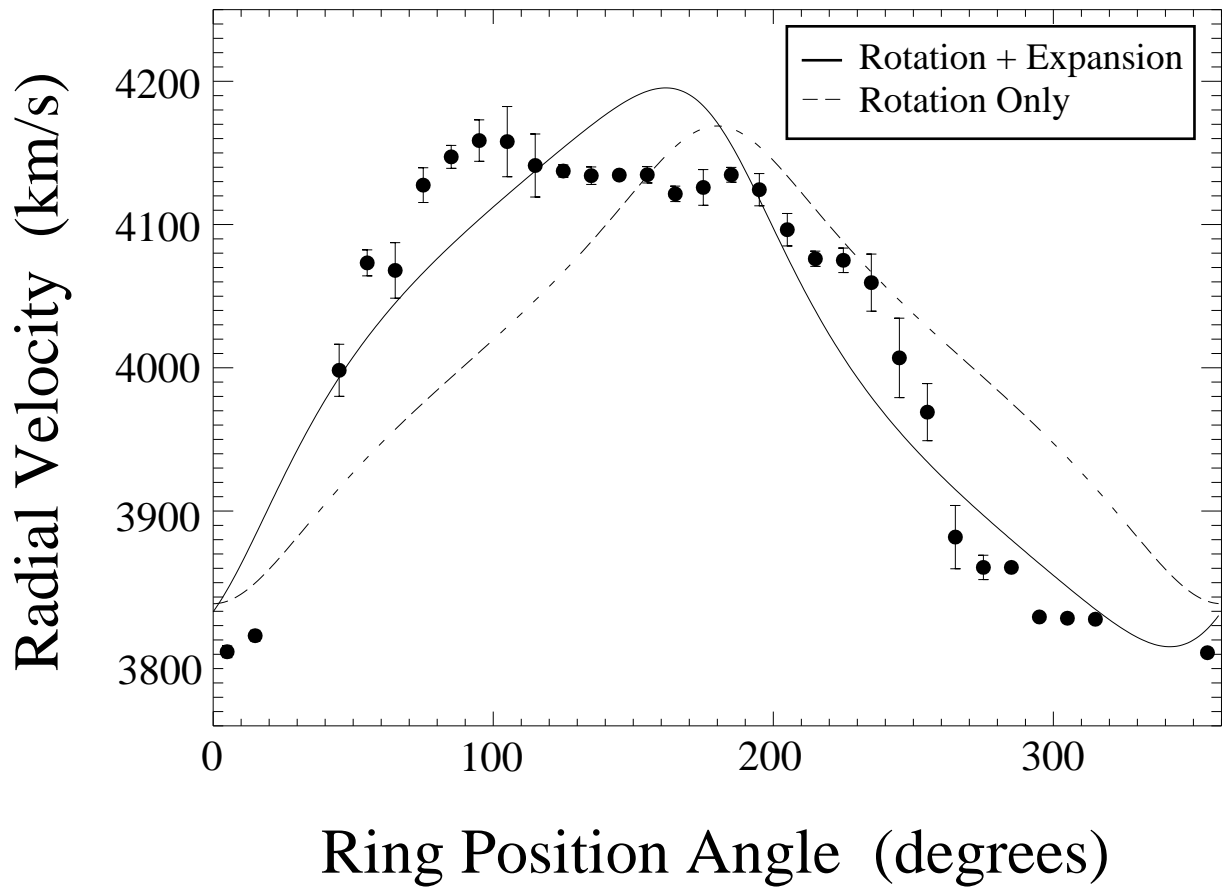
\end{document}